\documentclass[twoside,reqno]{bjp}
\usepackage{epsfig,cite}
\usepackage{graphics}
\usepackage{amssymb,amsmath}
\usepackage{times}
\begin{document}

\sloppy \raggedbottom

 \setcounter{page}{1}

\title{Tracing Quasar Accretion Rates at Higher Redshifts}

\runningheads{Quasar Accretion Rates}{Bachev}

\begin{start}
\author{R. S. Bachev}{1}

\address{Institute of Astronomy, Bulg. Acad. Sci., Sofia 1784, Bulgaria.\\ bachevr@astro.bas.bg}{1}

\received{}

\begin{Abstract}

In this work we connect some measurable properties of the CIV$\lambda$1549 emission 
line with the quasar accretion rates (Eddington ratios). A tight correlation is found 
for a sample of more than a hundred nearby objects, suggesting a possible method for a 
relatively accurate estimate of the Eddington ratio of high-redshift quasars, at least 
for the radio-quiet ones. This paper further confirms the existing notion that the CIV 
changes (shifts) are mostly driven by the accretion rate.

\end{Abstract}

\PACS {....}
\end{start}

\section[]{Introduction}

Quasars are long regarded as one of the most powerful objects in the Universe. 
It was not until recently, when it was realized the existence of a deep connection between 
the evolution of the quasars and the evolution of their host galaxies (e.g. Magorrian et al. 1998). 
Therefore, studying quasars is not only important for better understanding the physics involved 
there, taking into account that too many aspects still remain unclear, but also for better 
understanding the evolution of the galaxies, starting from its earliest stages.

Since the beginning of the SDSS project, the number of known quasars has increased rapidly. 
Most of these objects, however, are located at high redshifts and their traditionally most 
studied spectral region -- the optics -- is not easily accessible from the ground. The optical 
emission lines (especially H$\beta$) and luminosities were mostly used to estimate some of the quasars' 
defining characteristics, such as the black hole masses and the accretion rates (Eddington ratios) 
either by direct or indirect methods (Peterson \& Horne 2004; Kaspi et al. 2005). On the other hand, 
the higher redshift reveals the UV region of the quasar spectra and makes accessible some strong, 
important UV lines, such as CIV$\lambda$1549, among others. There have been attempts to substitute H$\beta$ 
with CIV in the methods used for mass estimates (Vestergaard 2002), but the results did not seem to be 
convincing enough, especially taking into account the different behavior of these two lines, the 
latter of which often shows systemic blueshifts (Sulentic et al. 2000, for a review; Richards et al. 2002). 
Instead, it has been suggested that mostly the Eddington ratio, not the central mass, is what 
appears to drive the changes (shifts) of CIV (Bachev et al. 2004; Warner et al. 2004; Baskin \& Laor 2005).

In this work we derive an empirical relation between CIV profile measures (mostly related to the 
shift and the equivalent width, EW) and the Eddington ratio, computed following Kaspi et al. (2005) from the optical 
spectra (H$\beta$ region). For this purpose we used a sample of 124 quasars with good quality optical spectra 
(Marziani et al. 2003) and also with good UV spectra (HST archive), covering the CIV region. 
The simultaneous coverage of the optical and the UV regions allowed searching for a relation between 
the quasar parameters (mass, accretion rate) and different CIV profile measures.

\section{Results}
\begin{figure}
\centering{\epsfig{file=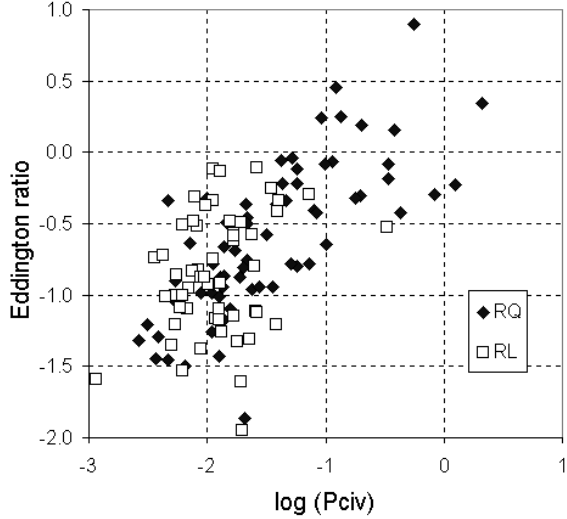,width=80mm}}
\caption{The relation between log($L/L_{\rm edd}$) and log($P_{\rm CIV}$), where $P_{\rm CIV}$ is 
the EW-normalized ratio of the blueward and redward profile widths at half maximum; see the text. 
124 objects are included, 75 of which are radio-quiet. The radio-quiet objects are shown as filled 
rhombs, the radio-loud ones -- as open squares. Note that RQ objects show a much better correlation, 
as well as that only they show apparent super Eddington accretion rates.}
\end{figure}

Exploring different quantities, we found a parameter ($P_{\rm CIV}$), related to the EW-normalized shift of CIV, 
which correlates very well with the Eddington ratio ($L/L_{\rm edd}$). This parameter is the ratio of the 
blueward and redward (measured in respect to the systemic velocity) widths of the profile at half maximum, 
normalized by the EW (in Angstroms) of the CIV line, i.e. 
$P_{\rm CIV}=(FW_{\rm blue}/FW_{\rm red})/EW$. This parameter should not be difficult to measure, 
provided a spectrum of good quality is available and the precise redshift of the object is known. 
We found a very good correlation between $L/L_{\rm edd}$ and $P_{\rm CIV}$ (in logarithmic units), 
with a Pearson correlation coefficient of 0.64 (Fig. 1). The correlation even improves further 
if only radio-quiet objects are considered, 0.73, and slightly more if $P_{\rm CIV}$ is measured 
at zero-intensity instead of half-maximum level. Thus, one can use $P_{\rm CIV}$ to 
estimate $L/L_{\rm edd}$, following the derived empirical relation (bisector linear fit):


$${\rm log}(L/L_{\rm edd}) = 0.64~{\rm log}(P_{\rm CIV}) - 0.47$$


with a typical error of 0.5 dex for $L/L_{\rm edd}$ (Fig. 1). One is to note, however, that $P_{\rm CIV}$ 
applies to the broad component of CIV (after the subtraction of a NLR contribution on the top, if such is 
present, Bachev et al. 2004) and also that this parameter is calculated upon the redshift based on 
[OIII]$\lambda$5007 or the narrow top of H$\beta$ lines, not the top of CIV (often blueshifted).
It should also be mentioned, that this relation is found mostly for powerful quasars and may not 
hold for the lower-luminosity objects (i.e. for $L/L_{\rm edd} < 0.01$; Fig. 1).

\section{Discussion and Conclusions}

The exact nature of correlations like the one we demonstrated above is not known, but one may suspect 
it has something to do with the accretion disk winds. As it is often assumed, CIV has a wind origin 
(Murray et al. 1995; Proga \& Kallman  2004) and therefore it is not surprising to find a connection 
between the line profile and the Eddington ratio, taking into account that the later is often thought 
to drive the wind strength (details in Bachev et al. 2004, and the references within). Why exactly 
$P_{\rm CIV}$, as defined, appears however to be the best surrogate for the Eddington ratio, should be a matter 
of future modeling. We also note that the "$P_{\rm CIV}$ -- $L/L_{\rm edd}$" relation could be of importance 
not only for modeling CIV emission region, but also for explaining the radio-quiet/radio-loud differences 
(e.g. Fig. 1). From a practical point of view, one can use such empirical relations for roughly estimating the 
Eddington ratio at high redshifts, where no other estimators are available. Thus, the early accretion 
history of the quasars can be traced, provided CIV properties do not change significantly with cosmic time.

\end{document}